\documentclass[pra,twocolumn,showpacs,preprintnumbers]{revtex4}
\usepackage{graphicx}
\usepackage{amsfonts}
\usepackage{amssymb}
\usepackage{amsmath}
\makeindex
\newcommand{\be}{\begin{equation}}
\newcommand{\ee}{\end{equation}}
\newcommand{\bea}{\begin{eqnarray}}
\newcommand{\eea}{\end{eqnarray}}
\newcommand{\Eq}[1]{Eq.\,(\ref{#1})}
\newcommand{\Fig}[1]{Fig.\,\ref{#1}}
\newcommand{\Sec}[1]{Sec.\,\ref{#1}}
\newcommand{\Onlinecite}[1]{Ref.\,\onlinecite{#1}} 

\newcommand{\GFk}{\hat{\mathbf{G}}_k}

\newcommand{\br}{\mathbf{r}}

\newcommand{\bE}{\mathbf{E}}

\newcommand{\En}{\mathbf{E}_n}
\newcommand{\Em}{\mathbf{E}_{m}}

\newcommand{\heps}{\hat{\boldsymbol{\varepsilon}}}
\newcommand{\hsigma}{\hat{\boldsymbol{\sigma}}}

\newcommand{\epsBK}{\epsilon_{\rm g}}
\newcommand{\sigmaBK}{\sigma_{\rm g}}

\newcommand{\epsSI}{\epsilon_{\rm Si}}
\newcommand{\sigmaSI}{\sigma_{\rm Si}}

\begin{document}
\title{Resonant-state expansion for a simple dispersive medium}
\author{ M.\,B. Doost}
\author{W. Langbein}
\author{E.\,A. Muljarov}
\affiliation{School of Physics and Astronomy, Cardiff University, Cardiff CF24 3AA,
United Kingdom}
\begin{abstract}
The resonant-state expansion (RSE), a rigorous perturbative method developed in electrodynamics for non-dispersive optical systems is applied to media with an Ohm's law dispersion, in which the frequency dependent part of the permittivity scales inversely with the frequency, corresponding to a frequency-independent conductivity. This dispersion has only a single pole at zero frequency, which is already present in the non-dispersive RSE, allowing to maintain not only the linearity of the eigenvalue problem of the RSE but also its size. Media which can be described by this dispersion over the relevant frequency range, such as optical glass or doped semiconductors, can be treated in the RSE without additional complexity. Results are presented using analytically solvable homogeneous spheres, for doped silicon and BK7 glass, both for a perturbation of the system going from non-dispersive to dispersive media and the reverse, from dispersive to non-dispersive media. %

\end{abstract}
\date{\today}
\maketitle
\section{Introduction}
The electromagnetic spectrum of an open optical system is characterized by its resonances, having spectral positions and linewidths,  corresponding to, respectively, the real ($\Omega$) and imaginary ($\Gamma$) part of the complex eigenfrequencies $\omega=\Omega-i\Gamma$ of the system. Finite linewidths ($2\Gamma$ full width at half maximum) of resonances are typical for open systems due to energy leakage from the system to the outside, and are described by the quality factor $Q=\Omega/2\Gamma$. While the eigenmodes of resonators for a few highly symmetric geometries can be calculated exactly, simulating general system with high quality factors presents a significant challenge for the established simulation techniques such as the finite element method (FEM) or finite difference in time domain (FDTD) method. Recently, approximate approaches using resonant modes have been reported~\cite{TureciPRA06, RubinPRA10, WiersigPRA12, SauvanPRL13, VialPRA14}. We have developed~\cite{MuljarovEPL10} a rigorous perturbation theory called resonant-state expansion (RSE) and subsequently applied it to one-dimensional (1D), 2D and 3D systems~\cite{DoostPRA12,DoostPRA13,ArmitagePRA14,DoostPRA14}, demonstrating its ability to accurately and efficiently calculate resonant states (RSs) -- the eigenmodes of a perturbed open optical system, -- using the spectrum of RSs of a simpler, unperturbed one.

Up to now, the RSE assumed a frequency independent permittivity. This is a significant limitation for the treatment of realistic systems which are all exhibiting dispersion to some extend. In this paper we extend the RSE formulation from non-dispersive media to media exhibiting an Ohm's law dispersion, i.e. having a permittivity consisting of a constant part plus a frequency dependent part scaling inversely with the frequency, as would be expected for a frequency-independent conductivity. This dispersion has only a single pole at zero frequency, which is already present in the non-dispersive RSE, allowing to maintain the linearity of the matrix eigenvalue problem and its dimension. This can be understood as due to the property of this conductivity not to add memory to the system, and thus not to introduce additional material resonances a finite frequency which would act as an additional degree of freedom. Apart from conductive materials also other dispersive materials such as optical glass can be described by this permittivity over a relevant frequency range using an imaginary conductivity, and therefore can be treated in the RSE without additional complexity.

The paper is organized as follows. \Sec{sec:RSE} outlines the RSE with an Ohm's law dispersion. Examples illustrating the method and comparing results with existing analytic solutions are given in \Sec{sec:application3D}. Specifically, we show in \Sec{sec:Si} the doping of a silica sphere, and in \Sec{sec:BK7} an application to a dispersive glass sphere.
\section{RSE with Ohm's law dispersion}\label{sec:RSE}
Resonant states of an open optical system with a local time-independent relative permittivity (RP) tensor $\heps_k(\br)$ and relative permeability $\mu=1$ are defined as the eigen-solutions of Maxwell's wave equation,
\be
\label{me3D}
\nabla\times\bigl[\nabla\times\En(\br)\bigr]=k_n^2\heps_{k_n}(\br)\En(\br)\,,
\ee
satisfying the {\it outgoing wave} boundary conditions. Here, $k_n=\omega_n/c$ is the wave-vector eigenvalue of the RS numbered by the index $n$, $\omega_n$ is its complex eigenfrequency, and $c$ is the speed of light in vacuum.  In the present work, we consider the RP
\be \heps_k(\br)=\heps(\br)+\dfrac{i\hsigma(\br)}{k}, \label{eqn:eps} \ee
which includes additionally to the frequency-independent term $\heps(\br)$, a term with the conductivity $\hsigma(\br)$, in accordance with Ohm's law.

A perturbed optical system is described by the wave equation
\be \label{me3Dpert}
\nabla\times\bigl[\nabla\times\bE(\br)\bigr]=k^2\bigl[\heps_k(\br)+\Delta\heps_k(\br )\bigr]\bE(\br)\ee
with
\be \Delta\heps_k(\br)=\Delta\heps(\br)+\frac{i\Delta\hsigma(\br)}{k}\,,
\label{del-eps}
\ee
in which the RS wave vectors $k$ and the electric fields $\bE$ are modified as compared to $k_n$ and $\En$, respectively, due to a perturbation $\Delta\heps_k(\br)$ inside the system. We treat this problem {\em exactly} by using the Green's function (GF) $\GFk(\br,\br')$ of the unperturbed Maxwell's wave equation satisfying the outgoing wave boundary conditions, which yields a formal solution of \Eq{me3Dpert}:
\be \bE(\br)=-k^2 \int \GFk(\br,\br')\Delta\heps_{k}(\br')\bE(\br') d{\bf r}'\,.
\label{GFsol} \ee
Substituting the spectral representation of the GF~\cite{DoostPRA13}, 
\be \GFk(\br,\br')=\sum _n\frac{\En(\br)\otimes\En(\br')}{2 k_n(k-k_n)}\,,
\label{ML1}\ee
into the unperturbed Maxwell's wave equation with a source term $\hat{\mathbf{1}}\delta(\br-\br')$
we obtain~\cite{MuljarovARX14} the closure relation
\be \dfrac{\heps(\br)}{2}\sum_n\En(\br)\otimes\En(\br')=\hat{\mathbf{
1}}\delta(\br-\br') \label{Closure} \ee
and the sum rule 
\be \sum_n \dfrac{\En(\br)\otimes\En(\br')}{k_n}=0\,,
\label{Sum-rule} \ee
both having the same form as in the non-dispersive case~\cite{DoostPRA13}.
Combining \Eq{ML1} and \Eq{Sum-rule} yields~\cite{DoostPRA13}
\be \GFk(\br,\br')=\sum_n \frac{\En(\br)\otimes\En(\br')}{2k(k-k_n)}\,.
\label{ML4} \ee
This representation of the GF is then used in \Eq{GFsol} leading to the following integral equation
\be
\label{GFexpansion}
\bE(\br) = \sum_n\frac{\En(\br)}{2(k_n-k)}\int\En(\br')\cdot k\Delta\heps_k(\br') \bE(\br')d{\bf r}'\,.
\ee
Expanding the perturbed wave functions into the unperturbed ones,
\be \bE(\br) = \sum _n c_{n} \En(\br)\,,
\label{expansion}
\ee
and using the linear independence of the basis functions $\En(\br)$, which holds for any finite basis size, we equate for each $n$ the expansion coefficients at $\En(\br)$ in both sides of \Eq{GFexpansion}. For the perturbation in the form of \Eq{del-eps}, this results in a linear generalized matrix eigenvalue problem
\be \sum_{m}(k V_{nm}+iS_{nm})c_m=2(k_n-k)c_n\,,
\label{matr}
\ee
in which
\bea V_{nm}&=&\int \En(\br)\cdot\Delta\heps(\br)\Em(\br)\,d \br \label{Vnm}\,,\\  S_{nm}&=&\int \En(\br)\cdot\Delta\hsigma(\br)\Em(\br)\,d \br\,. \label{Snm} \eea
Without perturbation in the conductivity,  $S_{nm}$ vanish, and \Eq{matr} reduces to the previously known matrix equation of the RSE, which can be reduced to a symmetric matrix eigenvalue problem [see Eqs.\,(14) and (15) in \Onlinecite{MuljarovEPL10}].
All modes appearing in Eqs.\,(\ref{ML4})--(\ref{Snm}) are normalized \cite{MuljarovARX14} according to
\bea
\label{normaliz}
1&=&\int_V\En(\br)\cdot\dfrac{\partial(k^2\heps_k(\br))}{\partial(k^2)}\biggr\rvert_{k=k_n}\!\!\En(\br)d{\bf r}\\
&&+\frac{1}{2k^2_n}\oint_{S_V} \left(\En\cdot\frac{\partial}{\partial
s}r\frac{\partial\En}{\partial r}-r\frac{\partial \En}{\partial r} \cdot \frac{\partial \En}{\partial
s}\right) dS\nonumber
\eea
where the first integral is taken over an arbitrary simply connected volume $V$ enclosing the inhomogeneity of the system and the center of the spherical coordinates used, and the second integral is taken over its surface $S_V$ with $\partial/\partial s$ being the
gradient normal to the surface. This normalization is required \cite{DoostPRA14} for the validity of the spectral representation \Eq{ML4}.

\section{Application to spherical 3D systems with scalar permittivity}\label{sec:application3D}

In this section we show applications of the simple dispersive RSE to 3D systems described by a scalar RP, i.e. $\heps_k(\br)=\hat{\mathbf{1}}\varepsilon_k(r)$, and $\Delta\heps_k(\br)=\hat{\mathbf{1}}\Delta\varepsilon_k(\br)$. As unperturbed system we use a homogeneous sphere of radius $R$ and $\epsilon_k =\epsilon+i\sigma/k$ in vacuum, so that
\be
\varepsilon_k(r)=\left\{
\begin{array}{lc}  \epsilon_k  & \mbox{for}\ \  r<R\,,\\
1 & \mbox{elsewhere}\,.\end{array}\right.
\ee

We treat transverse magnetic (TM) and transverse electric (TE) modes, while  longitudinal electric (LE) modes which all have $k=0$ are not present in the dispersive basis due to the divergence of $\epsilon_k$ at $k=0$ and vanishing normalization coefficients.

Owing to the spherical symmetry, the modes can be classified according to the orbital quantum number $l$  and magnetic quantum number $m$.
The modes $\En(\br)$ of this system are of the same analytical form as in \Onlinecite{DoostPRA14}. The resonant wave numbers $k_n$ are given by the solutions of known \cite{StrattonBook41} secular equations, see for example Eqs.\,(30) and (31) in \Onlinecite{DoostPRA14}. For a dispersive $\epsilon_k$ the normalization coefficients $A^{\rm TE}$ and $A^{\rm TM}$ given in \Onlinecite{DoostPRA14} are modified~\cite{MuljarovARX14} to
\be A^{\rm TE}_{l}=\sqrt{\frac{2}{l(l+1)R^3(\epsilon_k-1+\eta B_l)}} \,,
\label{ATE}
\ee
\be A^{\rm TM}_l=\frac{\displaystyle \sqrt{\frac{2\epsilon_k}{l(l+1)R^3(\epsilon_k-1)}}}{\displaystyle \sqrt{
\left(\frac{j_{l-1}(z)}{j_l(z)}-\frac{l}{z}\right)^2+\epsilon_k\frac{l(l+1)}{z^2}+
\eta C_l}}\,,
\label{A-normTM}\ee
where $z=\sqrt{\epsilon_k}k R$, and the additional terms in the normalization coefficients due to the dispersion are given by
\be\eta=\frac{1}{\epsilon_k}\frac{\partial(k^2 \epsilon_k)}{\partial(k^2)}-1\,, \ee
\be B_l=\epsilon_k\left(1-\frac{j_{l+1}(z)j_{l-1}(z)}{j_l^2(z)}\right)\,,\ee
\be C_l=\frac{\epsilon_k}{\epsilon_k-1} \left(2\frac{l\!+\!1}{z^2}+ \frac{j_{l+1}^2(z)}{j_l^2(z)}-\frac{j_{l+2}(z)}{j_l(z)}\right),
\label{Cl}
\ee
where all the quantities above are evaluated at $k=k_n$.

\subsection{Doping of a silicon sphere}\label{sec:Si}
\begin{figure}
\includegraphics*[width=\columnwidth]{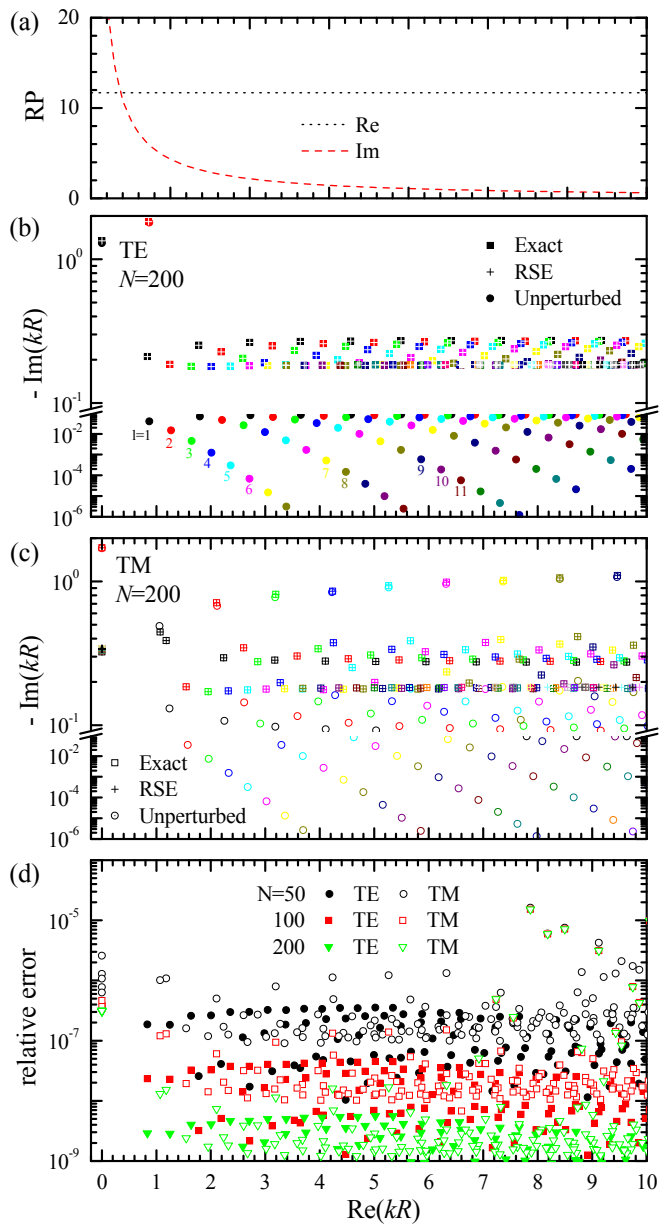}
\caption{
RSE results for the doping of a silicon sphere, as function of the real part of the dimensionless wave number $kR$. (a) Real and imaginary part of the RP $\epsilon_k$ of doped silicon. (b,c) RS wave numbers for TE (b) and TM (c) polarization and $N=200$. Shown are exact unperturbed (circles), exact perturbed (squares), and RSE-perturbed (crosses) data. The symbol color indicates $l$ as labeled. (d) Relative error $\delta$ of the RSE-perturbed RS wave numbers $k$ for $N=50,100$, and 200, for TE and TM modes with ${\rm Im}(kR)<1$, as labeled.}
  \label{fig:SiToDisp}
\end{figure}

Here we show as example a homogeneous $n$-doping of a silicon sphere at room temperature. The unperturbed system is a sphere of undoped silicon with $\epsilon=\epsSI=(3.42)^2$. The intrinsic conductivity of about $10^{-4}\,$S/m is neglected and we use $\sigma=0$. Introducing an $n$-doping with a density of $10^{14}/$cm$^3$ leads to a conductivity of $\sigmaSI=2.3\,$S/m \cite{WillisAPL13}. The Drude scattering time is about 200\,fs for this doping density \cite{WillisAPL13}, dominated by phonon scattering, such that the Ohm's law description is valid within 10\% error for frequencies below 200\,GHz, corresponding to $|kR| < 21$ for $R=5\,$mm. The resulting perturbation is $\Delta\sigma=\sigmaSI/(\epsilon_0 c)=4.334/R$ with the vacuum permittivity $\epsilon_0$, while $\Delta\epsilon=0$. The resulting RP of the doped silicon is shown in \Fig{fig:SiToDisp}(a) as function of $kR$.

For perturbations retaining the spherically symmetry, RSs of different angular quantum numbers $l$, $m$ and different transverse polarizations are not mixed, and are degenerate in $m$. The procedure used for the RSE calculation follows that described in \Onlinecite{DoostPRA14}. Namely, we choose the basis of RSs for the RSE in such a way that for given $l$, $m$ and TE or TM polarization, we select all RSs with  $|k_n|<k_{\rm max}(N)$ using a maximum wave vector $k_{\rm max}(N)$ chosen to select $N$ RSs.

The resulting RS wave vectors are given in \Fig{fig:SiToDisp}(b) for TE and \Fig{fig:SiToDisp}(c) for TM states, for the unperturbed system ($k_n$) and for the perturbed system ($k$). The perturbed wave vectors calculated using the RSE with $N=200$ are compared with the exact result $k^{\rm (exact)}$ from the secular equations.
The relative error $\delta=\bigl|{k}/{k^{\rm (exact)}}-1\bigr|$ is shown in \Fig{fig:SiToDisp}(d) for both TE and TM polarizations. We find that as we increase $N$, $\delta$ decreases proportional to $N^{-3}$, similar to the findings for the non-dispersive RSE \cite{DoostPRA14, DoostPRA13, DoostPRA12}, and values below $10^{-8}$ are reached for $N=200$.

Interestingly, the Ohms law dispersion gives rise to nearly constant imaginary part of the whispering gallery modes (WGMs) for large $kR$, which is due to the contribution of the absorption in the medium. In detail, $k=k_0/\sqrt{\epsilon_k} =k_0/\sqrt{\epsilon+i\sigma/k}\approx k_0/\sqrt{\epsilon}(1-i\sigma/(2 k\epsilon)$, for $|\sigma/(k\epsilon)|\ll 1$, where $k_0$ is the peak wave vector of the dissipated wavepacket, resonant to the frequency of the given RS: $k\approx k_0/\sqrt{\epsilon}$.
Then the imaginary part of the RS wavenumber is approximately $-\sigma/(2\epsilon)$,
which has a value of about $0.18/R$ for the doped silicon considered. This value is close to the imaginary part of WGMs seen in Figs.\,\ref{fig:SiToDisp}(b) and (c) for $kR>5$. The Fabry-Perot modes (FPMs) have additional losses due to the finite reflection at the silicon to vacuum interface, resulting in an additional imaginary part of about $0.09/R$. Leaky modes (LM) are mostly outside the sphere and are therefore less influenced by the absorption.

\begin{figure}
\includegraphics*[width=\columnwidth]{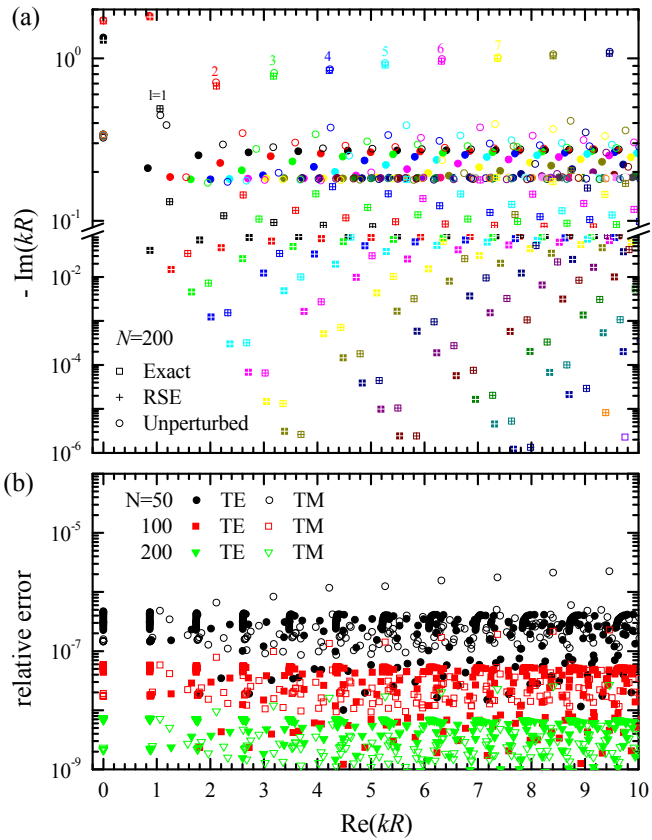}
\caption{
RSE results for the un-doping of a silicon sphere, as function of the real part of the dimensionless wave number $kR$. (a) RS wave numbers for both TE (filled symbols) and TM (open symbols) polarization and $N=200$. Shown are exact unperturbed (circles), exact perturbed (squares), and RSE-perturbed (crosses) data.  The symbol color indicates $l$ as in \Fig{fig:SiToDisp}. (b) Relative error $\delta$ of the RSE-perturbed RS wave numbers $k$ for $N=50,100$, and 200, for TE and TM modes with ${\rm Im}(kR)<1$, as labeled.}
  \label{fig:SiToNonDisp}
\end{figure}
We now exemplify that the RSE can also be used with a basis from a dispersive system, which is perturbed to a non-dispersive one. We use the $n$-doped silicon sphere as unperturbed system and perturb it by removing the doping, i.e. $\Delta\sigma=-4.334/R$. The resulting RS wave numbers and their relative error are given in \Fig{fig:SiToNonDisp} for different basis sizes $N$. We find that the wave numbers are well reproduced, and that the relative errors are scaling equivalently to the case of a non-dispersive basis.

\subsection{Dispersive glass sphere} \label{sec:BK7}

\begin{figure}
\includegraphics*[width=\columnwidth]{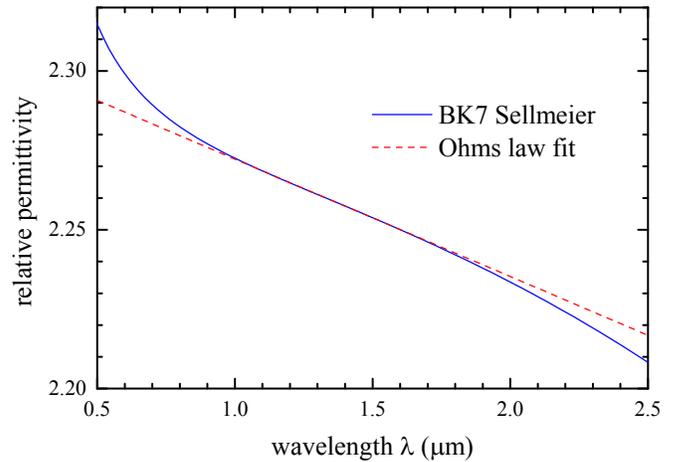}
\caption{RP $n_r^2$ of Schott BK7 glass (solid line), and fitted Ohm's law type dispersion $\epsilon_k$ (dashed line).}\label{fig:BK7Fit}
\end{figure}

The form of the dispersion, representing Ohm's law with a frequency independent conductivity, is specifically suited to describe conductors at frequencies below the charge carrier scattering rate. However, the mathematical treatment is in general allowing for complex $\epsilon$ and $\sigma$, which can be used to model other types of materials.
As example we use a technical glass in the optical frequency range, specifically the borosilicate glass Schott BK7, which has a refractive index $n_r$ in the optical frequency range described by the Sellmeier expression \cite{BK7Sellmeier}
\bea\label{eqn:BK7}
n_r^2&=&1+\frac{1.03961212\lambda^2}{\lambda^2-6000.69867\,\mbox{nm}^2}\\
&&+\frac{0.231792344\lambda^2}{\lambda^2-20017.9144\,\mbox{nm}^2}+
\frac{1.01046945\lambda^2}{\lambda^2-103.560653\,\mu\mbox{m}^2}\,.
\nonumber\eea
The resulting RP is real and is shown in \Fig{fig:BK7Fit}. It is fitted by $\epsilon_k$ using the parameters $\epsilon$ and $\sigma$ over the wavelength range from $1\,\mu$m to $1.8\,\mu$m. The resulting $\epsilon_k$, which is linear in wavelength $\lambda=2\pi/k$, is also shown in \Fig{fig:BK7Fit}, and is given by $\epsilon=\epsBK=2.30926$ and $\sigma=\sigmaBK=0.232414i/\mu$m. The deviation of the fit is below 0.001 over the fitted range. One has to keep in mind that a complex conductivity results in an unphysical dispersion, which does not have the property $\varepsilon_k=\varepsilon^\ast_{-k^\ast}$ following from the causality principle~\cite{LandauBook60}. The latter requires in particular that the conductivity has to be real. However, this does not affect the suitability of the RSE to describe the RSs close to the fitted $k$ range, and we  have evaluated the error of the fit on a selected segment of $k$. We note that the value of the permittivity at the generally complex $k_n$ will deviate to some extend from a physical RP which obeys Kramers-Kronig relationships.

\begin{figure}
\includegraphics*[width=\columnwidth]{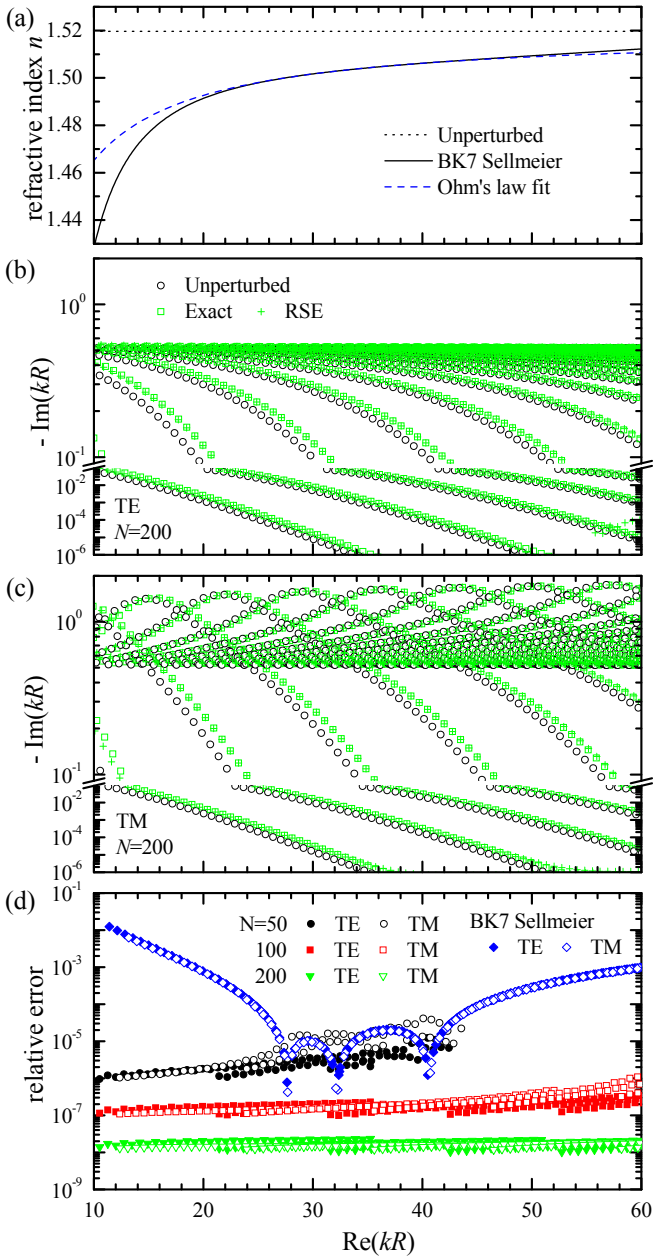}
\caption{RSE results for a homogeneous sphere with a non-dispersive unperturbed and a dispersive perturbed medium, as function of the real part of the dimensionles wave number $kR$. (a) Refractive index of the unperturbed (dotted line) and perturbed (dashed line) medium, together with the BK7 Sellmeier dispersion (solid line) given by \Eq{eqn:BK7}. (b,c) RS wave numbers for (TE,TM) polarization and $N=200$. Shown are exact unperturbed (open circles), exact perturbed (open squares), and RSE-perturbed (crosses) data. (d) Relative error $\delta$ of the RSE-perturbed RS wave numbers $k$ for $N=50,100$, and 200, for TE and TM modes with ${\rm Im}(kR)<0.1$, as labeled. For the $N=200$ data the relative error to the exact RS wave numbers for the Sellmeier  dispersion is also shown.
}\label{fig:BK7ToDisp}
\end{figure}

We first show RSE results introducing dispersion as perturbation of a non-dispersive system. The unperturbed system is given by a sphere with $\epsilon=\epsBK$, $\sigma=0$ and $R=7\,\mu$m. We consider a homogeneous perturbation over the sphere, having $\Delta\sigma=\sigmaBK$, $\Delta\epsilon=0$, so that the perturbed system has the fitted BK7 dispersion. We choose the basis of RSs for the RSE in the same way as in \Sec{sec:Si}. The results are given in \Fig{fig:BK7ToDisp}. In \Fig{fig:BK7ToDisp}(a) the refractive index of the unperturbed and perturbed system is shown as function of $kR$, and compared with the BK7 Sellmeier data. We can see that for the chosen sphere radius, the fitted wavelength range corresponds to $22<kR<55$.
The RS wave vectors are given in \Fig{fig:BK7ToDisp}(b) for TE and \Fig{fig:BK7ToDisp}(c) for TM states, for the unperturbed system ($k_n$) and for the perturbed system, and the relative error $\delta$ is shown in \Fig{fig:BK7ToDisp}(d). We again find that as we increase $N$, $\delta$ decreases proportional to $N^{-3}$, reaching values in the $10^{-8}$ range for $N=200$. This is actually smaller than the relative error due to the Ohm's law approximation of the Sellmeier dispersion, which is in the $10^{-4}-10^{-5}$ range, as also shown in \Fig{fig:BK7ToDisp}(d). For practical applications, the simple dispersive RSE is thus limited by the Ohm's law approximation of the dispersion over a given spectral range, but the achievable accuracy in the $10^{-4}$ range can be sufficient for many applications.
\begin{figure}
\includegraphics*[width=\columnwidth]{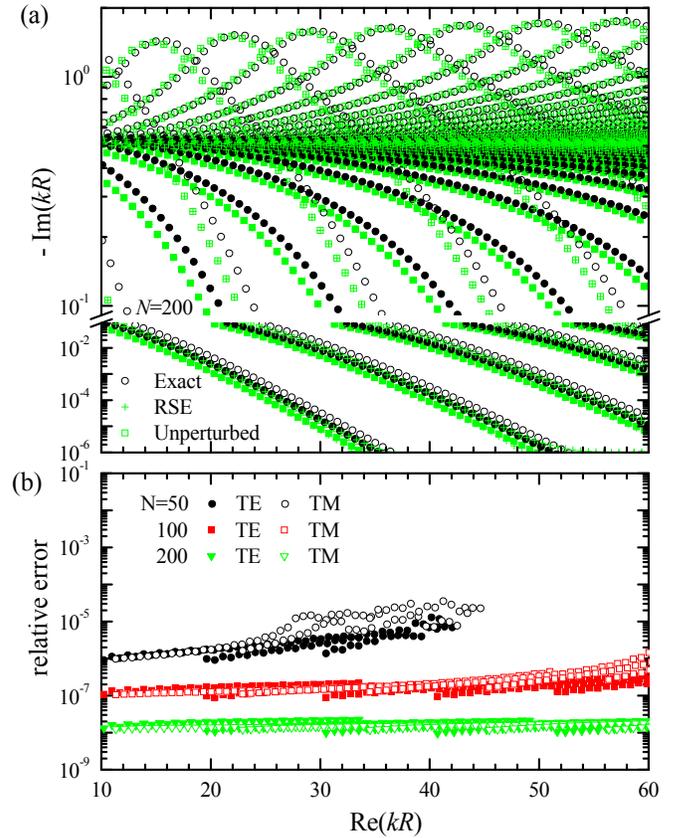}
\caption{
RSE results for a homogeneous sphere, with a dispersive unperturbed and a non-dispersive perturbed medium, as function of the real part of the dimensionless wave number $kR$. (a) RS wave numbers for both TE and TM polarization and $N=200$. Shown are exact unperturbed (open circles), exact perturbed (open squares), and RSE-perturbed (crosses) data. (b) Relative error $\delta$ of the RSE-perturbed RS wave numbers $k$ for $N=50,100$, and 200, for TE and TM modes with ${\rm Im}(kR)<0.1$ , as labeled.} \label{fig:BK7ToNonDisp}
\end{figure}

Also for this example we now verify that one can also use a dispersive basis to treat a dispersive perturbation. We revert the dispersive perturbation, so that the unperturbed system is given by a sphere with $\epsilon=\epsBK$, $\sigma=\sigmaBK$ and the perturbation is $\Delta\sigma=-\sigmaBK$, $\Delta\epsilon=0$. The resulting RS wave numbers and their relative error are given in \Fig{fig:BK7ToNonDisp} for different basis sizes $N$. We find that the wave numbers are well reproduced, and that the relative errors are scaling equivalently to the case of a non-dispersive basis shown in \Fig{fig:BK7ToDisp}.

\section{Summary}\label{sec:summary}
In this work we have extended the RSE to media having a simple dispersion equivalent to Ohm's law, i.e. a frequency independent conductivity. This dispersion has a single pole at zero frequency and does not introduce additional dynamic degrees of freedom. This property allows to keep the simplicity of the RSE formulation, with the only change to move from a normal matrix eigenvalue problem to a generalized matrix eigenvalue problem, keeping the matrix size the same. With suited algorithms, this results typically only in a doubling of the computational time, therefore retaining the advantage of the RSE in computational efficiency discussed in \Onlinecite{DoostPRA14}. The Ohm's law is a good approximation for conductive materials at frequencies well below the carrier scattering rate, which is typically in the terahertz frequency range at room temperature, and we show as example doped silicon. Furthermore, we have shown that by choosing an imaginary conductivity, this dispersion can also describe the dispersion of the real part of the permittivity of optical glass in the normal dispersion range, and the RSE is working equally well  despite the dispersion not respecting causality.

\acknowledgments M.D. acknowledges support by the Cardiff University EPSRC Doctoral Price Fellowship EP/L505390/1. E.A.M. and W.L. acknowledge support by the Cardiff University EPSRC Impact Acceleration Account EP/K503988/1, and S\^er Cymru National Research Network Cymru National Research Network in Advanced Engineering and Materials. E.A.M. acknowledges support from RFBR Grant No. 14-02-00778.

\end{document}